\documentclass[twoside]{lcnote}
\usepackage[latin1]{inputenc}
\usepackage[dvips]{graphicx,epsfig,color}
\usepackage{wrapfig,rotating}
\usepackage{amssymb,amsmath,array}
\usepackage[colorlinks,dvips,ps2pdf]{hyperref}

\begin{document}
\title{
ZVMST: a minimum spanning tree-based vertex finder} 
\author{S. Hillert
\vspace{.3cm}\\
University of Oxford - Department of Physics \\
Denys Wilkinson Building, Keble Road, Oxford OX1 3RH - UK
}

\maketitle

\begin{abstract}
A new topological vertex finder is presented which combines ideas of the well-established ZVTOP 
algorithm with a novel minimum spanning tree approach. A preliminary performance study with simulated 
$e^{+}e^{-} \rightarrow q\bar{q}$ events at a centre of mass energy of $\sqrt{s} = 92\, \mathrm{GeV}$ 
shows that the new approach is competitive with existing vertex finder algorithms.
\end{abstract}
\section{Introduction}
\label{sec:Introduction}
Minimum spanning trees (MSTs) are a mathematical optimisation tool with a wide range 
of applications, which include finding the best route to travel from one place to 
another, finding the optimal way to connect computers in a network and astrophysical 
applications, such as source detection in gamma ray images \cite{MSTForGammaRayDetection}, 
where the suggestion of using this approach for cluster finding dates back to the early 
1980s. 
As this method exploits topological information - in the astrophysical example the 
connectedness of the detected photons - one can expect this method to provide a natural 
approach to topological vertex finding, i.e. the identification of decay vertices of 
heavy flavour hadrons in the jets that result from the collision of high-energy 
particles at collider experiments.

For pixel-based vertex detectors of sufficient spatial resolution, the topological vertex finder 
ZVTOP, originally developed at the SLD experiment \cite{Jackson:1996sy} and more recently widely 
used in Linear Collider physics studies, see e.g. \cite{RDR:2007,Behnke:2001qq}, 
provides the most precise vertex information available to-date. 
It consists of two algorithms, ZVRES, which uses topological information to resolve secondary 
vertices from the primary in the dense track environment near the interaction point (IP), and ZVKIN, 
which is more specific in being only applicable to $b$-jets as it uses kinematic information only 
available for this flavour. 
In this paper, a new algorithm, ZVMST, is presented, which combines ideas of the ZVRES algorithm 
with a new ansatz for finding vertices based on a minimum spanning tree. 

The paper is structured as follows: after a brief introduction to minimum spanning trees in 
section \ref{subsec:MinSpanTree} and the mathematical functions used to describe the topological 
information which both ZVRES and ZVMST exploit in section \ref{subsec:GaussTubeAndVtxFct},
the new ZVMST algorithm is described in detail in section \ref{subsec:ZVMSTalgo}. 
The performance of the new algorithm is presented and compared to that of ZVRES and of a vertex 
cheater in section \ref{sec:Performance}. The vertex cheater is described in section 
\ref{subsec:VertexCheater}. 
The performance study comprises results on vertex multiplicity (section \ref{subsec:VertexMultiplicity}), 
purity of track-content of the vertices found (section \ref{subsec:TrackToVertexPurity}) and 
resulting flavour tagging performance (section \ref{subsec:FlavourTagPerformance}). 
Preliminary results regarding the dependence of some of these performance indicators on the parameters of the new algorithm are presented in section \ref{sec:ParameterChoice}.
Section \ref{sec:Conclusion} concludes with a summary and overview of further studies that would be 
useful to improve the understanding of the algorithm's performance and its optimal use for flavour 
tagging.
%
%
\section{The ZVMST vertexing algorithm}
\label{sec:ZVMST}
The new ZVMST algorithm is based on essentially the same mathematical description of 
the topological information of a given input jet that was carefully developed for ZVRES, 
and differs only in the way this information is used for finding vertices. 
The main challenge of vertex finders used prior to ZVTOP is the large number of track 
combinations that need to be considered and checked as to whether they form a good vertex. 
To avoid this problem, the ZVRES algorithm uses a bottom-up approach, starting out from 
all possible two-track combinations and using the vertex function as well as the 
fit-$\chi^{2}$ to decide which candidates to keep and to merge. 
While ZVRES is thus an iterative procedure that gradually arrives at optimised vertices, 
in the new ZVMST algorithm much of this optimisation is provided by the minimum spanning 
tree approach.

Before describing the new algorithm, it is therefore necessary to give a brief overview of 
minimum spanning trees and of the mathematical description of tracks and of the vertex 
function that is calculated from the track information.
%
\subsection{Minimum Spanning Trees}
\label{subsec:MinSpanTree}
Mathematically, minimum spanning trees are a special type of graph. A graph is a set of 
\textit{nodes} that are connected by \textit{edges}. Each edge can have a \textit{weight} 
assigned to it. For example, the nodes could correspond to different cities, the connecting 
edges to the roads linking these cities and the weights to the distances. (For route planning 
other criteria, such as the likelihood of traffic jams on particular roads, need to be 
taken into account to determine the weights).

\textit{Trees} are graphs in which for any pair of nodes there exists exactly one sequence of 
connecting edges - i.e. the edges do not form loops. For a graph containing loops, there 
therefore exist several trees, i.e. graphs with the same set of nodes as the original graph, 
but only a subset of the edges.

For graphs with weighted edges, the \textit{minimum spanning tree} is defined as the tree for 
which the subset of edges is chosen such as to minimise the sum of the weights. It can be 
shown that unless there are weights that are identical, there always exists exactly one 
minimum spanning tree for a given graph.

Efficient algorithms for finding the minimum spanning tree for a graph exist, e.g. the 
Dijkstra algorithm \cite{Dijkstra:1959} and the algorithm by Kruskal \cite{Kruskal:1956}. 
Well tested, optimised implementations of these algorithms are available in the graph 
library of the C++ package boost \cite{boost:2001}. This library has been used for the 
application to topological vertex finding.
%
\subsection{Track probability tubes and the vertex function}
\label{subsec:GaussTubeAndVtxFct}
A central idea of the ZVRES algorithm is to describe each track by a probability 
density function $f_{i}(\vec{r})$ in 3D space and to use these to define a vertex function 
$V(\vec{r})$ that yields higher values in the vicinity of true vertex locations and lower 
values elsewhere, as well as providing a criterion for when two vertex candidates 
are resolved from each other. 

The track functions have a Gaussian profile in the plane normal to the trajectory. 
With $\vec{p}$ the point of closest approach of track $i$ to space point $\vec{r}$ 
the track function $f_{i}(\vec{r})$ is defined as: 
\[ f_{i}(\vec{r}) = 
   \exp \left\{-\frac{1}{2} \left(\vec{r} - \vec{p}\right) {\mathbb{V}}_{i}^{-1}
                            {\left(\vec{r} - \vec{p}\right)}^{T}
        \right\} \ ,
\]
where ${\mathbb{V}}_{i}$ is the covariance of the track at $\vec{p}$. 

In its most basic form, the vertex function is then defined as 
\[ V(\vec{r}) = \sum_{i=1}^{N}{f_{i}(\vec{r})} - 
                \frac{\sum_{i=1}^{N}{f_{i}^{2}(\vec{r})}}{\sum_{i=1}^{N}{f_{i}(\vec{r})}}
\]
with the second term preventing a single track passing near a point from yielding a high 
$V(\vec{r})$ value at that position.
Optionally, further knowledge on where vertices are more likely to be found can be used 
to weight the vertex function, thereby suppressing fake vertices and increasing the purity 
of the vertices found (i.e. the fraction of correctly assigned tracks). 
Knowledge about the location of the event vertex can be used to suppress fake vertices from 
tracks passing close by each other in the vicinity of the IP location. 
By adding a contribution 
\[ f_{0}(\vec{r}) = 
   \exp \left\{-\frac{1}{2} \left(\vec{r} - \vec{p}\right) {\mathbb{V}}_{IP}^{-1}
                            {\left(\vec{r} - \vec{p}\right)}^{T}
        \right\}\ ,
\]
with $\mathbb{V}_{IP}$ representing the IP covariance and $\vec{p}$ the IP position, 
and redefining the vertex function as 
\[ V(\vec{r}) = f_{0}(\vec{r}) + \sum_{i=1}^{N}{f_{i}(\vec{r})} - 
                \frac{f_{0}^{2}(\vec{r}) + \sum_{i=1}^{N}{f_{i}^{2}(\vec{r})}}
		     {f_{0}(\vec{r}) \sum_{i=1}^{N}{f_{i}(\vec{r})}}\ ,
\]
space points close to the IP are less likely to be resolved from each other 
and tracks that could otherwise give rise to fake vertices are more likely to be 
assigned to the primary vertex. This IP-contribution is not adopted for the new ZVMST 
algorithm, as it can also have the side-effect of suppressing secondary vertices that 
could otherwise be found.
	
Similarly, vertices are more likely to be found close to the jet axis than at a large 
angle from it, which is taken into account by re-weighting the vertex function outside 
a cylinder of radius $50\,\mu m$ by an attenuation factor 
$ \exp \left(-K_{\alpha} \alpha^{2} \right)$ 
with opening angle $\alpha$, 
$ K_{\alpha} = k E_{\mathrm{Jet}}$
with $k$ a user-settable code parameter and $E_{\mathrm{Jet}}$ the jet energy. 
This jet-energy dependent definition of $K_{\alpha}$ takes into account that jets of higher 
energy are more collimated.

In addition to indicating likely vertex positions, the other use of the vertex function 
in the ZVRES algorithm is to provide a key criterion for merging candidate vertices in 
the process of vertex finding: space points $\vec{r_{1}}$ and $\vec{r_{2}}$ are defined to 
be resolved from each other, if along the straight line connecting these points the vertex 
function falls below a given fraction $R_{0}$ of the lower of the values $V(\vec{r_{1}})$ 
and $V(\vec{r_{2}})$. 
%
\subsection{The ZVMST vertex finder}
\label{subsec:ZVMSTalgo}
The ZVMST algorithm has two main stages: first a small number - typically between 1 and 5 - 
of 3D positions at which vertices are likely to be found is chosen on the basis of the 
vertex function. In the second phase tracks are assigned to these candidate vertex positions, 
using both the value of the Gaussian probability tube of each track at each of the selected 
space points and the vertex function value at these points. 

To select the candidate vertex positions, the intial step is identical with that of ZVRES: 
for all possible two-track combinations in the input jet a vertex-fit is attempted, and 
combinations discarded that have a fit-$\chi^{2}$ above a user-settable cut value (default: 
10) or for which the vertex function at the resulting fit position is below 0.0001. 
In contrast to ZVRES, combinations of one track and the IP are not considered in this
approach. \footnote{Including these combinations has been tried and results in some non-primary
vertices not being found, as the one-track IP combinations tend to have large vertex
function values (regardless of goodness of fit) and hence tend to be preferably selected
in the following minimum spanning tree based optimisation step, leading to non-primary
vertex positions being discarded and some vertices not being found.} 

The retained two-track combinations are used to set up a mathematical graph structure, in 
which each node corresponds to one of the tracks in the jet, and each edge corresponds to a 
successful vertex fit of the two tracks that it connects. Note that a connection is only made 
if the corresponding fit passes the cuts described above. As weight for the edge, the inverse 
of the vertex function at the vertex position obtained from the two-track fit is chosen. 

The graph is passed as input to the minimum spanning tree algorithm. This algorithm selects a 
set of at most $N-1$ edges for $N$ input nodes (or less if the input graph contains unconnected 
nodes) in such a way that the overall weight is minimised. In this case, because of the choice 
of the weights, this minimisation corresponds to maximising the sum of the vertex function 
values for the selected two-track candidate vertices. 

Often some of the $N-1$ selected candidates will correspond to the same physical vertex, 
especially for multi-prong vertices and the primary vertex. Therefore, sets of two-track 
candidates that correspond to one physical vertex need to be identified and only one optimised 
position derived for each set. One can expect that two-track candidate vertices corresponding 
to the same physical vertex should have fit-positions that are close in space and thus also 
have similar vertex function values. This information should therefore be useful in deciding, 
whether two candidate vertices are representing the same physical vertex or not. Due to measurement 
uncertainties, the fit-position obtained from the two-track fit may be shifted with respect to 
the position one would obtain from a simultaneous fit of all tracks belonging to the physical 
vertex. Better results are therefore obtained if one searches the vicinity of each two-track 
candidate for the 3D position which maximises the vertex function. The "maximal vertex function 
value in the vicinity of a candidate vertex" is also used in the later stages of ZVRES, and the 
same implementation is used for the new algorithm.

For ZVMST, the sorted list of $N-1$ two-track fits that results from the minimum spanning tree 
algorithm is used to select candidate 3D positions as follows: an empty list of candidate 
positions is created. For each of the $N-1$ two-track fits it is checked, if its maximal vertex 
function position corresponds to the same physical vertex as any of the candidate positions in 
the list. Only if none of the positions fulfils this criterion, the maximal vertex function 
position of the two-track fit is added to the list of candidate positions. Two positions are 
considered to correspond to the same physical vertex if their spatial separation is below a 
user-settable cut value $d_{\mathrm{min}}$ (default: $400\,\mu m$). A more detailed discussion 
of this criterion, and possible further studies related to it can be found in section 
\ref{sec:Conclusion}.

The first stage of the ZVMST algorithm is now complete, and has resulted in a short list of 
candidate 3D positions, which are considered in the second stage of the algorithm, concerned 
with assigning tracks to vertices.

For each track $i$ the probability tube value is calculated for each position in the list of 
candidate 3D positions. The two largest of these for each track, 
$f_{i}\left(\vec{r}_{i1}\right)$ and $f_{i}\left(\vec{r}_{i2}\right)$, with 
$f_{i}\left(\vec{r}_{i1}\right) > f_{i}\left(\vec{r}_{i2}\right)$
and the vertex function values for the two positions, $V(\vec{r}_{i1})$, $V(\vec{r}_{i2})$, 
are used to decide, if to assign track $i$ to candidate position $\vec{r}_{i1}$, to 
$\vec{r}_{i2}$ or not at all: 
for assigning a track to position $\vec{r}$ it is required that $f_{i}(\vec{r})$ is larger 
than $f_{\mathrm{min}}$ (default 0.0001). 
If both the track function and the vertex function are larger for $\vec{r}_{i1}$, the track 
is assigned to $\vec{r}_{i1}$.

However, there are also some tracks, for which $V(\vec{r}_{i2}) > V(\vec{r}_{i1})$ and the 
track functions are sizeable at both $\vec{r}_{i1}$ and $\vec{r}_{i2}$, and in these cases it 
is useful to consider assigning the track to the candidate position with larger vertex function, 
even if the track function at that position is smaller. 
If $V(\vec{r}_{i2}) > V(\vec{r}_{i1})$, it depends on the relative differences 
\[ \Delta f_{i1,2} = \frac{\left|f_{i}\left(\vec{r}_{i1}\right) - f_{i}\left(\vec{r}_{i2}\right)\right|}
                          {f_{i}\left(\vec{r}_{i1}\right) + f_{i}\left(\vec{r}_{i2}\right)} 
   \ \ \ \mathrm{and}\ \ \
   \Delta V_{i1,2} = \frac{\left|V_{i}\left(\vec{r}_{i1}\right) - V_{i}\left(\vec{r}_{i2}\right)\right|}
                          {V_{i}\left(\vec{r}_{i1}\right) + V_{i}\left(\vec{r}_{i2}\right)}\ ,
\]
if track $i$ is assigned to $\vec{r}_{i1}$ or $\vec{r}_{i2}$. 
The track is assigned to $\vec{r}_{i2}$, if the difference between the track 
functions at $\vec{r}_{i1}$ and $\vec{r}_{i2}$ is "not too large" and the vertex function 
difference is "large enough", i.e. if $\Delta f_{i1,2} < \Delta f_{\mathrm{max}}$ (default value: 
0.95) and $\Delta V_{i1,2} > \Delta V_{\mathrm{min}}$ (default: 0.15). 

The intial values for these cuts were chosen "by hand" based on these values found for number of 
example jets and the corresponding track-to-vertex correspondence known from the Monte Carlo (MC) 
generator, and subsequently an initial optimisation study was performed, which is described in 
section \ref{sec:ParameterChoice}. 
Effectively, these cuts do come close to the criterion of assigning the track not to the position 
with higher track function value, but to the one with higher vertex function value (as could have 
been expected from the role the vertex function plays for the track-to-vertex assignment in ZVRES). 
It was therefore also tried, for cases for which both $f_{i}\left(\vec{r}_{i1}\right)$ and 
$f_{i}\left(\vec{r}_{i2}\right)$ are above the minimum track function cut, to base the decision 
between $\vec{r}_{i1}$ and $\vec{r}_{i2}$ exclusively on the vertex function. 
However this resulted in less good performance than the cuts described above. 
%
%
\section{Performance of the ZVMST algorithm}
\label{sec:Performance}
The goal of vertexing algorithms is to find both the positions of the vertices in a jet and 
the set of decay tracks originating at each of these vertices. Despite this clear goal, there 
is no single performance measure for vertexing algorithms, but a range of criteria that should be 
taken into account. In the initial performance study presented in this note, the emphasis is 
on whether ZVMST can compete with existing algorithms in terms of finding vertices without 
creating an increased number of fakes, e.g. by wrongly combining two IP-tracks to form a 
non-physical secondary. In addition to a vertex multiplicity study, it is investigated to which 
extent the correct tracks are assigned to the found vertices. It can be expected, that if this 
aspect of vertexing is competitive, the resulting vertex positions should on average also be at 
least as good as those obtained from existing algorithms if the vertex fitting procedure is 
unchanged. The quality of the reconstructed vertex positions is not yet considered in the present 
study and will need to be evaluated in the future. Instead, it is investigated, how the performance 
of the identification of different jet-flavours (the flavour-tag) depends on the vertex finder 
used. For each of these aspects, the result obtained from ZVMST is compared to that of the 
well-established ZVRES algorithm, as well as a "vertex cheater" algorithm described in section 
\ref{subsec:VertexCheater} .

For the intial performance study presented in this note, a sample of 10000 two-jet 
$e^{+}e^{-} \rightarrow q\bar{q}$
events ($q\bar{q} = b\bar{b}, c\bar{c}, s\bar{s}, u\bar{u}, d\bar{d}$), generated with the PYTHIA 
event generator at a centre of mass energy of $\sqrt{s} = 92\,\mathrm{ GeV}$ was used.
The response of a typical ILC detector design, the LDC detector model \verb|LDC01_05Sc|, was 
simulated using the GEANT4-based program MOKKA. 
Events were reconstructed using the MarlinReco event reconstruction package. The recently 
developed LCFIVertex package provides a set of algorithms to perform topological vertexing 
(ZVRES and ZVKIN), flavour tagging and vertex charge reconstruction. The 
ZVMST algorithm was implemented into the LCFIVertex package, using the minimum spanning tree 
implementation from the graph library of the boost C++ library. The boost library is already 
being used by the LCFIVertex package for vector and matrix representation.  
The LCFIVertex code was also used for obtaining the flavour tag and track-to-vertex 
association purity. 
For the ZVMST algorithm, the default code parameters given in section \ref{sec:ParameterChoice} 
were used. 
%
\subsection{The vertex cheater algorithm}
\label{subsec:VertexCheater}
The idea of the vertex cheater algorithm is to provide a performance comparison with vertices 
obtained from perfect track-to-vertex assignment based on MC-truth information. The sets of 
tracks obtained in this way are passed to the same vertex-fitter used for reconstructed vertices. 
In this way the aspect of finding the correct track-combinations is disentangled from the problem 
of finding the correct vertex position. While the vertex positions found by the cheater deviate 
from the true MC vertex positions, this approach has the advantage that the vertices obtained from 
the cheater can be treated in the same way as vertices obtained from the topological vertex 
reconstruction algorithms, e.g. the fitter provides realistic covariance matrices, which are 
subsequently used in the calculation of the input variables from which the flavour tag is 
obtained.

It should be noted, that "perfect" track-to-vertex assignment is to be understood within the 
limitations imposed by performing vertex finding on a jet-by-jet basis, i.e. if the jet-finder 
assigns two tracks originating from a common decay vertex to different jets, this vertex will 
not be found by the cheater algorithm.
%
\subsection{Comparison of vertex multiplicity}
\label{subsec:VertexMultiplicity}
%
\begin{figure*}[h!]
\centering
\begin{tabular}{cc}
\mbox{\includegraphics[width=0.50\columnwidth]{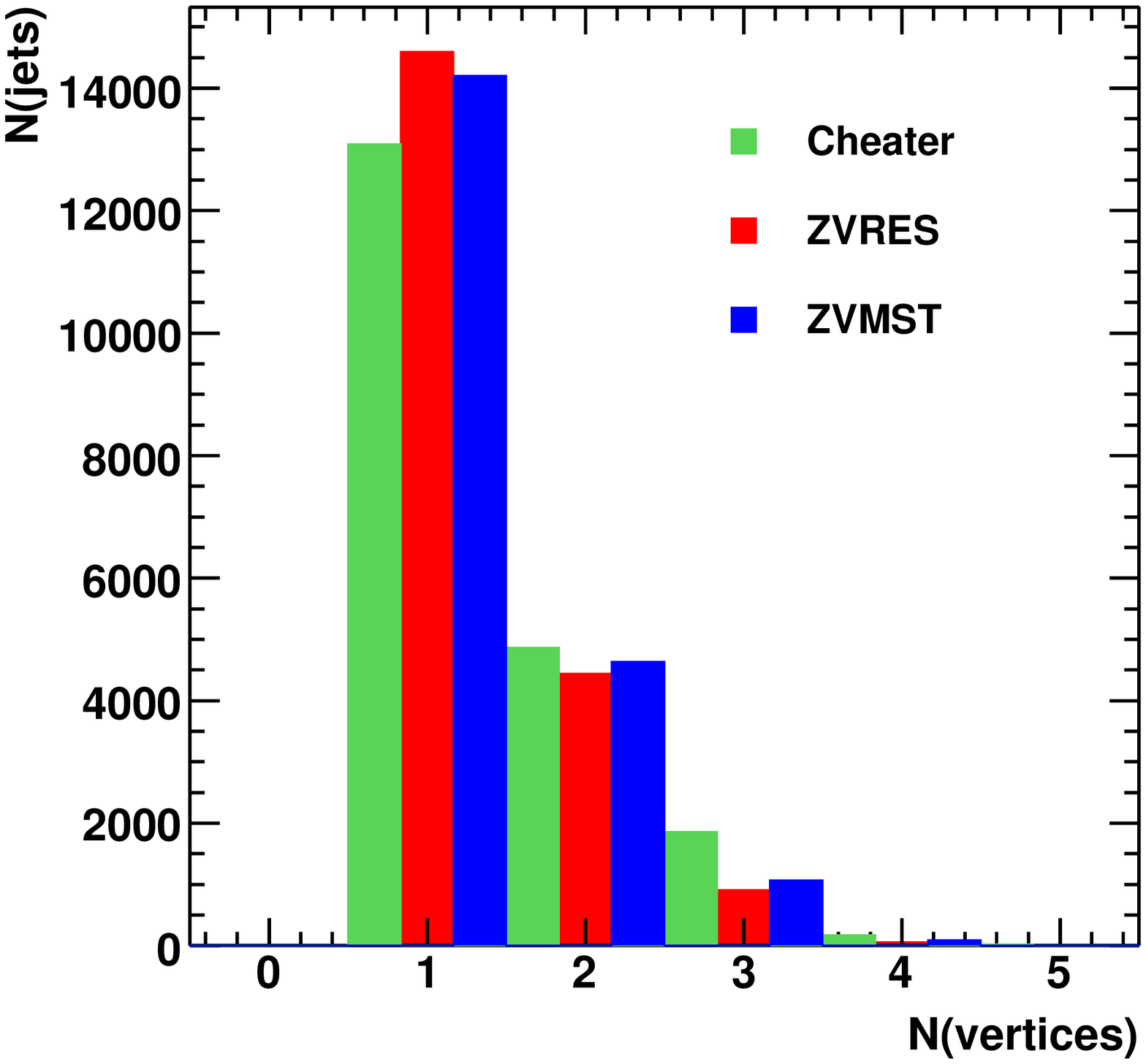}} &
\mbox{\includegraphics[width=0.50\columnwidth]{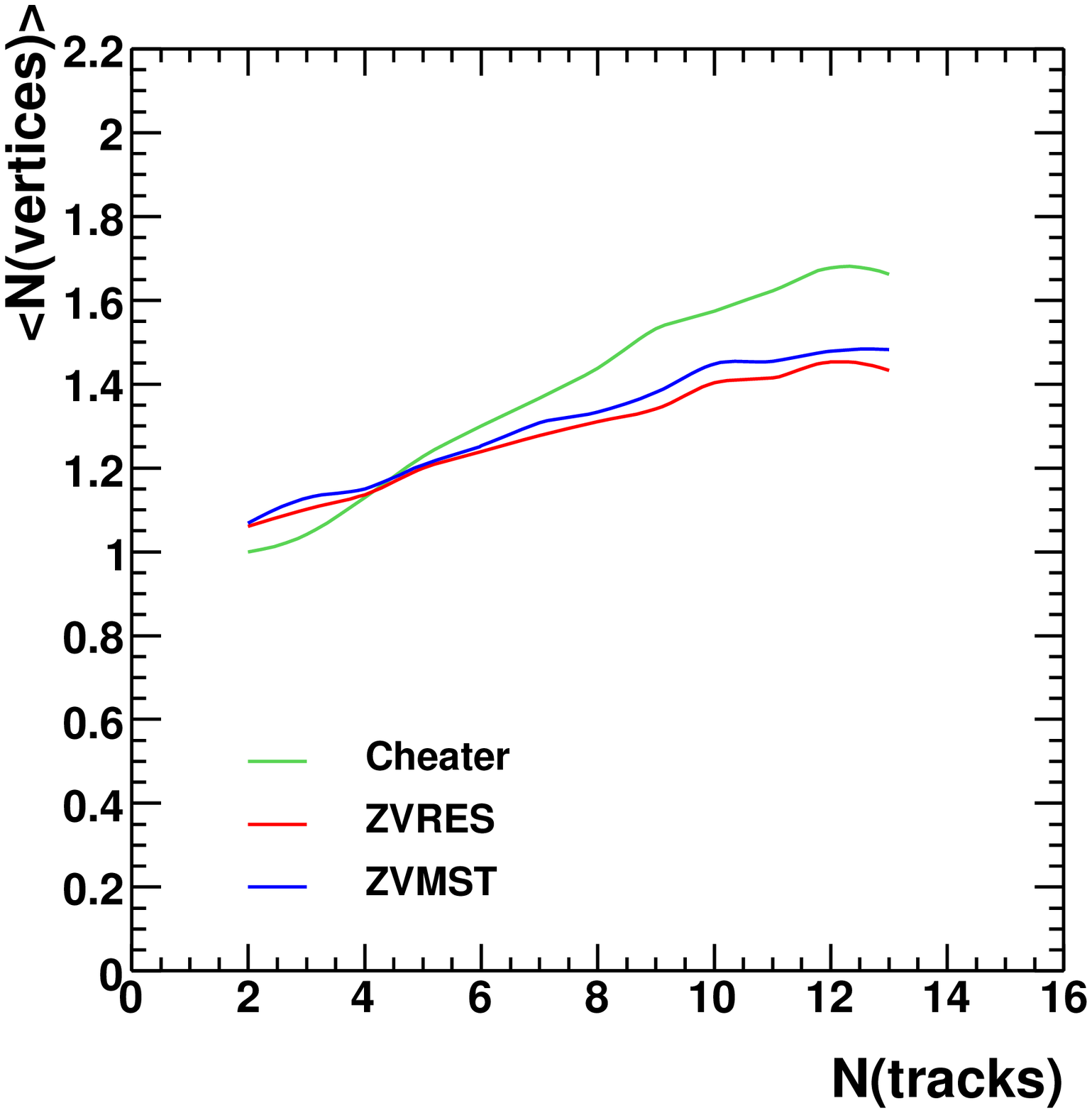}}\\
(a) & (b)\\
\end{tabular}
\caption{\textit{Multiplicity of vertices found by the two topological vertex reconstruction
algorithms ZVRES and ZVMST. The multiplicity of reconstructable vertices as found by
the vertex cheater is shown for comparison. (Reconstructable vertices are all vertices
that contain at least two tracks that are assigned to the same jet by the jet-finder).
Shown are (a) the inclusive distribution and (b) the average vertex multiplicity as 
function of track multiplicity in the input jet.}}
\label{Fig:VertexMultiplicity}
\end{figure*}
In Fig.~\ref{Fig:VertexMultiplicity} the multiplicity of vertices reconstructed by the 
topological vertex finders is compared to the one obtained from the vertex cheater. 
The inclusive distributions in Fig.~\ref{Fig:VertexMultiplicity} (a) show that as expected 
both the ZVRES and the ZVMST vertex finder reconstruct a smaller number of vertices than is 
actually present in the jets. 
It is known for ZVRES that this is especially the case at small decay lengths, where the 
danger of creating fake vertices is particularly high. The two reconstruction algorithms 
ZVRES and ZVMST yield very similar results, with the overall number of vertices found by 
ZVMST being slightly larger than for ZVRES. 

The dependence of vertex multiplicity on the number of input tracks in the jet is plotted 
in  Fig.~\ref{Fig:VertexMultiplicity} (b). The reconstructed vertex multiplicity 
is slightly higher for ZVMST than it is for ZVRES over the entire range of input track 
multiplicities. The comparison with the vertex cheater shows that while for high track 
multiplicity the tendency increases not to find some of the physical vertices that are present 
in the jet, for a track multiplicity of up to 4 both ZVRES and ZVMST find too many (i.e. fake) 
vertices.

This suggests that it might be useful to investigate if vertex finding could be improved 
by making the criteria for deciding whether to create more or fewer vertex candidates (the 
resolvability cut for ZVRES and the $d_\mathrm{min}$ cut for ZVMST) track-multiplicity 
dependent. Care would have to be taken not to introduce an unwanted dependence on jet-energy 
as a side-effect.
%
\subsection{Purity of track-content of reconstructed vertices}
\label{subsec:TrackToVertexPurity}
The extent to which the reconstructed vertices contain the correct tracks is quantified in 
terms of the purity of the track-to-vertex assignment as determined by the 
\verb|LCFIAIDAPlot-|\newline\verb|Processor| 
of the LCFIVertex package. As different types of jets present different challenges to vertex 
finding, different classes of jets and vertices are studied separately. In particular, separate 
studies are performed for the case that two vertices and that three vertices have been 
reconstructed in a jet, and for $b$-jets and $c$-jets. (Note that this is the MC-truth flavour, 
as determined from an angular match between the jet axis and the heaviest-flavour MC-hadron that 
is the direct parent of any of the MC-particles corresponding to the tracks in the jet, as 
implemented in the \verb|TrueAngularJetFinder| of LCFIVertex). For each of these four classes, 
the primary, secondary, and where available the tertiary vertex were studied separately. 
For each, the fractions of tracks were determined, for which the corresponding MC particle 
originated from the IP, the $B$-hadron decay and the $D$-hadron decay. 
\begin{table}{}
\centerline{\begin{tabular}{||cc|ccc|cccc||}
\hline
\multicolumn{2}{||c|}{Monte Carlo} &
\multicolumn{7}{c||}{Reconstructed track-vertex association} \\
\multicolumn{2}{||c|}{track origin} &
\multicolumn{3}{c}{Two vertex case ($b$)} &
\multicolumn{4}{c||}{Three vertex case ($b$)} \\
\hline
 & & \ \ pri\ \  & sec & iso & \ \ pri\ \  & sec & ter & iso \\
\hline
        & Cheater  & 93.8 & 0.508 & 24.2 & 98.7 & 0.167 & 1.29 & 46.8 \\
Primary & ZVRES    & 89.8 & 1.83  & 34.3 & 94.4 & 5.85  & 5.74 & 47.5 \\
        & ZVMST    & 89.1 & 2.31  & 46.6 & 95.5 & 5.78  & 5.85 & 57.7 \\
\hline
          & Cheater & 5.95 & 41.2 & 41.6 & 1.24 & 79.8 & 14.5 & 28.8 \\
$B$ decay & ZVRES   & 7.26 & 48.3 & 30.5 & 3.83 & 65.5 & 12.9 & 25.6 \\
          & ZVMST   & 8.05 & 48.5 & 22.6 & 3    & 67.3 & 17.4 & 19.7 \\
\hline
          & Cheater  & 0.242 & 58.3 & 34.2 & 0.0295 & 20   & 84.2 & 24.4 \\
$D$ decay & ZVRES    & 2.92  & 49.8 & 35.2 & 1.74   & 28.7 & 81.3 & 26.9 \\
          & ZVMST    & 2.81  & 49.2 & 30.8 & 1.53   & 27   & 76.7 & 22.5 \\
\hline
          & Cheater  & 52.4 & 31.1 & 16.5 & 40.7 & 28.8 & 22.7 & 7.8 \\
all above & ZVRES    & 49.3 & 36.9 & 13.8 & 38.2 & 29.7 & 23.4 & 8.73 \\
          & ZVMST    & 45.5 & 35.3 & 19.2 & 34.2 & 28.3 & 23.7 & 13.7 \\
\hline
\multicolumn{2}{||c|}{track origin} &
\multicolumn{3}{c}{Two vertex case ($c$)} &
\multicolumn{4}{c||}{Three vertex case ($c$)} \\
\hline
 & & \ \ pri\ \  & sec & iso & \ \ pri\ \  & sec & ter & iso \\
\hline
          & Cheater  & 99.8 & 0.862 & 74.4 & 99.9 & 7.16 & 26.7 & 75.9 \\
Primary   & ZVRES    & 94.8 & 7.17  & 75   & 94.3 & 26.7 & 33.6 & 67.4 \\
          & ZVMST    & 95.8 & 9.2   & 78.2 & 95.7 & 21.2 & 44.5 & 73.8 \\
\hline
          & Cheater  & 0.191 & 99.1 & 25.6 & 0.104 & 92.8 & 73.3 & 24.1 \\
$D$ decay & ZVRES    & 5.23  & 92.8 & 25   & 5.67  & 73.3 & 66.4 & 32.6 \\
          & ZVMST    & 4.18  & 90.8 & 21.8 & 4.29  & 78.8 & 55.5 & 26.2 \\
\hline
          & Cheater  & 65.4 & 27.2 & 7.44 & 56.3 & 19.7 & 18.9 & 5.11 \\
all above & ZVRES    & 64.1 & 27.3 & 8.59 & 49.5 & 22.3 & 21.1 & 7.1 \\
          & ZVMST    & 60.9 & 27.7 & 11.4 & 44.9 & 23   & 20.1 & 11.9 \\
\hline
\end{tabular}}
\caption{\textit{Percentages of the tracks assigned to the reconstructed primary,
secondary and tertiary vertex, which originate from the IP-, the $B$ or the
$D$ decay at MC level. The fractions of tracks assigned to primary, secondary
and tertiary vertex, normalised to the total number of tracks in the jet are
also shown. Four different types of jet, with $b$- and $c$-flavour and with
2 or 3 vertices found, are considered separately.}}
\label{table:TrackToVertexPurity}
\end{table}
The resulting purities obtained from the vertex cheater, ZVRES and ZVMST are shown in table 
\ref{table:TrackToVertexPurity}, along with the fractions of tracks contained in each type 
of vertex, normalised to the total number of tracks in each of the four jet categories.

The residual amount of confusion found for the vertex cheater despite the use of MC information 
for the track assignment indicates some of the limitations of this approach of studying 
track-to-vertex association purity, which should also be kept in mind when looking at the results 
from the reconstruction algorithms. These limitations include the following effects: 
\begin{itemize}
\item Non-primary one-prong vertices cannot be found by any of the three approaches, including 
the vertex cheater. 
For the case of the cheater, this is intentional, so the purities obtained this way can be used as 
reference for the two topological reconstruction algorithms. (Note that the ZVKIN reconstruction 
algorithm has the possibility to reconstruct such vertices; for studies of the ZVKIN performance, 
a modified vertex cheater taking this into account, could therefore be useful in future).
\item Which reconstructed vertex is considered the primary, the secondary and the tertiary is decided 
only on the basis of their distance from the IP/event vertex. \footnote{Before ZVTOP is run, the event 
vertex is reconstructed by an independent processor taking all tracks in the event as input.} 
This assignment may not be correct, i.e. particularly in low-energy jets there can be cases, for 
which the $D$-hadron emerges from the $B$-hadron decay vertex at angles larger than $90^{\circ}$ to the 
jet axis, resulting in the $D$ decay vertex being closer to the primary than the $B$ decay vertex.  
\item Hadronic interactions in the detector material give rise to additional vertices. These are 
currently not identified and known not have a significant effect on the flavour tag at the jet 
energy considered in this study. However, where present they do invalidate the assumption made 
regarding which reconstructed vertex corresponds to the decay of which type of MC-hadron.
\item Due to effects of pattern recognition at the track reconstruction stage, the correspondence 
between tracks and MC-particles is not perfect.
\end{itemize}

Bearing in mind these sources for the confusion inherent in the way the purities are determined, 
the vertex cheater provides the best purity values achievable, which can serve as a reference to 
which to compare the results from the two vertex reconstruction algorithms. This is particularly 
useful for the cases, in which the reconstructed number of vertices does not agree with what 
would be expected for the jet-flavour (i.e. 2-vertex $b$-jets and 3-vertex $c$-jets), and where it 
is therefore otherwise not clear, e.g. what fraction of the tracks in the secondary vertex of a 
$b$-jet with only two vertices found should be expected to come from the $B$- and from the $D$-decay, 
respectively.

For the three vertex case, the track-content of the primary and the secondary vertex is improved 
by ZVMST compared to the ZVRES result for both $b$- and $c$-jets. The improvement is particularly 
large for the secondary vertex in $c$-jets with three vertices, reducing the fraction of IP-tracks 
that are wrongly assigned to the secondary vertex by almost 1/5 as compared to the ZVRES result. 

However, for the tertiary vertex, ZVMST does not reach the performance of ZVRES in assigning the 
correct tracks, but shows an increased level of confusion with the primary vertex for $c$-jets, 
and between $B$- and $D$-decay tracks for $b$-jets.
	
For $b$-jets in which only two vertices are reconstructed, the confusion between primary, $B$- and 
$D$-decay tracks is larger for ZVMST than it is for ZVRES. For two-vertex $c$-jets, the reconstructed 
primary vertex contains a smaller fraction of tracks actually originating from the $D$-decay, but 
the fraction of IP-tracks assigned to the reconstructed secondary vertex is increased.

Regarding the distribution of tracks between the different types of vertices, the new ZVMST 
algorithm tends to assign less tracks than ZVRES, returning a larger fraction of isolated tracks. 
This fact might to some extent account for the improved purities found for ZVMST in some cases. 
To fully assess the track-assignment performance, the fraction of jets that fall into the four 
categories ($b$-, $c$- with 2-, 3-vertices) would need to be included in the study. 

In conclusion, the differences between the two topological vertexing algorithms are smaller than 
the differences between these reconstruction algorithms and the vertex cheater, indicating that 
rather different vertex reconstruction algorithms yield similar performance. What is less clear 
is how to decide on the basis of the track-to-vertex purities as discussed in this section, which 
of the two reconstruction algorithms performs better. This is partly due to the fact that the 
definition of the different classes of jets that are considered separately is not optimal: the 
same jet can fall into the 2-vertex category for ZVRES and the 3-vertex category for ZVMST or 
vice versa. It is therefore unclear whether to interpret an increased purity for one type of 
vertex as improved performance, as the correlations with the changes in purity for the other types 
of vertices and the fraction of isolated tracks are not known.

Complementary to the vertex purity results presented in table \ref{table:TrackToVertexPurity}, 
it might therefore be useful to explore different ways of assessing vertexing performance that 
would avoid these problems. For example, for each track that is assigned to a vertex by a 
reconstruction algorithm one could define $\Delta R_\mathrm{rec}$ as distance between the 
location of this reconstructed vertex and the MC origin of the particle corresponding to the track. 
One could then plot the fraction of tracks in bins of $\Delta R_\mathrm{rec}$. As isolated tracks 
would be included in the normalisation, there would be no doubt about whether apparent improvements 
are only due to less tracks having been assigned as is the case with the current assessment method.
%
\subsection{Flavour tag results}
\label{subsec:FlavourTagPerformance}
Finally, the performance of ZVMST is studied in terms of the resulting flavour tag purities 
for $b$-jets, $c$-jets and $c$-jets with $b$-background only. For each of these, the flavour 
tag is obtained from the output of a neural net fed with characteristic jet properties, such 
as the $P_t$-corrected vertex mass and the joint probability of the tracks to originate from 
a common vertex. Depending on whether one-, two- or more vertices are found, different neural 
nets are used. The flavour tagging approach was originally developed by R.~Hawkings 
\cite{Xella Hansen:2000dr}. The implementation of this approach provided by the 
\verb|FlavourTagInputsProcessor| and the \verb|FlavourTag| processor of the LCFIVertex 
package is described in detail elsewhere \cite{FutureNIMpaper:LCFIVertex}. 
For the study presented in this note, the neural nets that were trained with input from the 
fast MC SGV were used. 

\begin{figure*}[h]
\centering
\mbox{
\includegraphics[width=0.95\columnwidth]{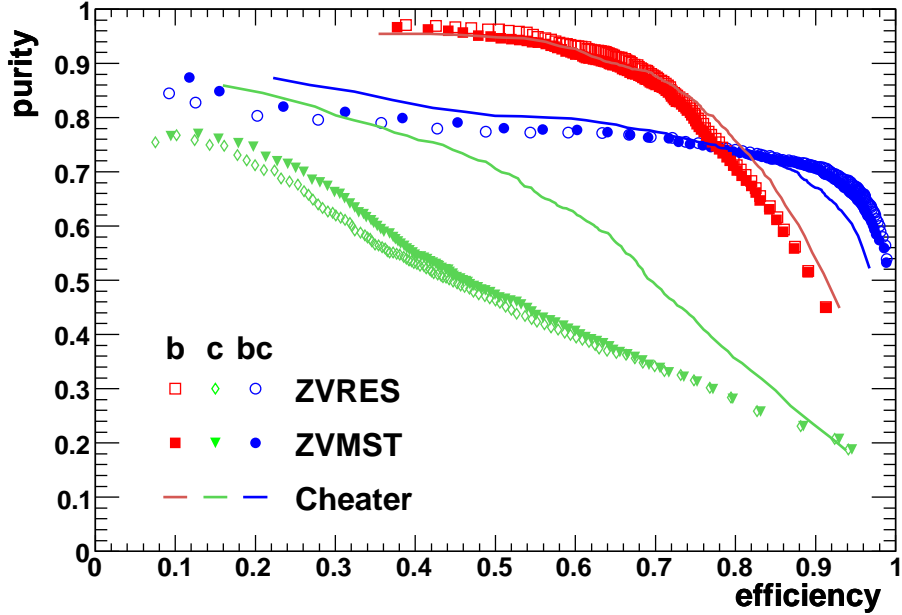}}
\caption{\textit{Comparison of tagging performance at the $Z$-resonance obtained using the 
new ZVMST vertex finder compared to results obtained using ZVTOP's ZVRES algorithm 
and for a vertex cheater using MC information for track-to-vertex assignment. 
Tagging purity is shown as function of efficiency for $b$-jets 
and $c$-jets. Performance for $c$-jets assuming only $b$-background (labelled ``bc'') is also 
shown. }}
\label{Fig:FlavourTagPerformance}
\end{figure*}

Fig.~\ref{Fig:FlavourTagPerformance} shows the resulting purities for the three tags as function 
of tagging efficiency for ZVMST. 
Results obtained from ZVRES and from the vertex cheater are shown for comparison. Again, 
performance obtained from ZVMST is similar to the one resulting from ZVRES. For the $c$-tag with 
all backgrounds, ZVMST performs better over the full $c$-tag efficiency range, at low efficiency 
giving an improvement of up to $5 \%$. In contrast, the $b$-tag purity is lower for ZVMST than 
it is for ZVRES, by up to $\approx1.5 \%$, which is probably related to the lower purity of 
the $c$-tag with $b$-background at high efficiency (with improvement wrt ZVRES for this tag at low 
efficiency). 

The cheater result for the $c$-tag shows that if further improvements in the track-to-vertex 
assignment could be made, this should directly result in corresponding improvements in the flavour 
tag. In contrast, the $b$-tag performance of both reconstruction algorithms is close to the result 
provided by the vertex cheater, with ZVRES yielding a higher $b$-tag purity than the cheater at low 
efficiency, indicating that the superior track-to-vertex assignment of the cheater for this tag 
does not improve the tagging performance when using the current set of flavour tag inputs and 
networks. 
The reason for this could be that the additional vertices that the cheater provides compared to 
ZVRES, mostly correspond to low decay lengths and have different characteristics than the vertices 
with which the flavour tag networks used in this study were trained. To some extent this could 
also be the case for ZVMST. For understanding the difference in $b$-tag performance between 
ZVRES and ZVMST, a study of the decay length dependence of vertexing performance of the two 
algorithms could therefore be useful.

Finally, the result obtained from the vertex cheater should not be misinterpreted as indicating 
the optimum flavour tagging performance achievable: There could be ways of improving the flavour 
tag by means other than an improvement in the track-to-vertex assignment, e.g. by including 
additional neural net input variables providing a different type of further information. Such 
information could include characteristics of the minimum spanning tree of the ZVMST algorithm, 
e.g. the number of edges and candidate 3D positions considered or the sum of the vertex function 
value corresponding to all selected edges.
%
%
\section{Choice of code parameters}
\label{sec:ParameterChoice}
\begin{table}{}
\centerline{\begin{tabular}{|c|c|c|c|c|}
\hline
parameter & initial value & range investigated & step size & resulting default value\\
\hline
$d_{\mathrm{min}}$ [$\mu m$] & 300 & [50, 600] & 50 & 400 \\
\hline
$f_{\mathrm{min}}$ & 0.0001 & $\left[0.5 \cdot 10^{-4}, 1.5 \cdot 10^{-4} \right]$ & $0.1 \cdot 10^{-4}$ & 0.0001 \\
\hline
$\Delta f_{\mathrm{max}}$ & 0.9 & [0.85, 0.95] & 0.05 & 0.95 \\
\hline
$\Delta V_{\mathrm{min}}$ & 0.1 &[0.05, 0.15]  & 0.01 & 0.15 \\
\hline
\end{tabular}}
\caption{\textit{Parameters of the ZVMST algorithm included in the preliminary study of 
parameter dependence. The ranges over which each parameter was varied and the 
step size by which parameters were incremented in these intervals are shown, 
as well as the current default values derived from the study.}}
\label{table:ParameterTuning}
\end{table}
Compared to ZVRES, the ZVMST algorithm has five new parameters, as described in detail in 
section \ref{subsec:ZVMSTalgo}. Initial values for these parameters were selected on the 
basis of printout of values for 20 randomly picked events, with the true origin of the 
MC particle corresponding to each track being available for comparison. A preliminary tuning 
of these parameters was then performed, in which each of the five parameters was varied over 
a range of values while keeping the other four parameters fixed. The initial default values, 
ranges of variation and resulting defaults (quoted in section \ref{subsec:ZVMSTalgo}) are 
shown in table \ref{table:ParameterTuning}. 
The optimised values were again 
selected "by hand" on the basis of the flavour tag purities at 70, 80 and $90\%$ tagging efficiency. 
This choice was not always unambiguous, as sometimes different values were favoured by the different 
tags and the different efficiencies. Generally, a very weak parameter dependence was seen over 
most of the parameter ranges considered for all the parameters. Further work will be required to 
understand the parameter dependence in more detail.
%
%
%
\section{Summary and Conclusion}
\label{sec:Conclusion}
A new topological vertexing algorithm based on a minimum spanning tree approach, ZVMST, 
was presented and shown to yield results that are competitive with the leading existing 
algorithm ZVRES. 

Vertex multiplicities for the new approach are slightly closer than ZVRES to the reference 
results that were obtained by using MC information. A study of dependence of reconstructed 
vertex multiplicity on the input track multiplicity suggests that an improvement might be 
possible by taking the track multiplicity into account in deciding how tight to choose the 
cuts that determine the multiplicity of candidate vertices. Care will need to be taken not 
to accidentally introduce a jet-energy dependence as a side-effect. 

Both algorithms yield similar results when studying track-to-vertex assignment. The current 
method of assessing this aspect of vertex finding was shown to have some limitations making 
it difficult to interpret the results. A different approach was suggested that would also 
take the distance between reconstructed and MC vertices into account. 

In terms of flavour tagging performance, the new algorithm yields an improvement of up to 
$5\%$ in purity for the $c$-tag, while performing slightly less well than ZVRES in terms of 
$b$-tag purity. The flavour tagging performance yielded by both algorithms is very similar,
while differing considerably from the results obtained from the vertex cheater.Future 
improvements of the neural net-based flavour tag might be possible by adding further 
input variables, which could include characteristics of the minimum spanning tree found 
for each jet by the ZVMST algorithm.

It should be noted that the results presented are very preliminary. In particular, the 
algorithm parameters of both ZVRES and ZVMST are not yet optimised. Also, some of the choices 
made during the development of ZVMST should be reconsidered. In particular the criterion for 
deciding whether two candidate vertex positions correspond to the same physical vertex will 
need to be revisited. Alternatively to the distance cut used in the current approach, the 
criterion of when two vertices are resolved from each other as implemented in ZVRES could be 
used. Both criteria (distance and resolvability) have been tried during the development phase, 
with the distance criterion yielding slightly better results for the $c$-tag. However, since 
both criteria depend on a cut value, which for the ZVRES criterion was kept fixed, this 
comparison is largely inconclusive. A systematic study of a range of $R_{0}$ and $d_\mathrm{min}$, 
for a range of jet energies would be required to decide which of the two criteria to prefer. 
For example, the resolver criterion might prove to be less sensitive to jet energy. On the 
other hand, should there be a jet-energy dependence of the optimal $d_\mathrm{min}$ value, 
it might be possible to account for this by a simple scaling.

Further studies could also show, if the two algorithms have particular strengths for a 
subset of jets with characteristics that allow to decide which vertex finder to use. 


\begin{footnotesize}

\end{footnotesize}


\end{document}